\documentstyle[eqsecnum,revtex]{aps}

\begin{document}
\draft
%**************************************************
 
\begin{title}
 PROBLEMS WITH THE STANDARD SEMICLASSICAL
  \\
 IMPACT LINE-BROADENING THEORY \\

\end{title}

\author{Spiros Alexiou}
\begin{instit}
 P2IM, URA 773, Case 232, Universite de Provence, Centre St.Jerome
 13397 Marseille, France\\
\end{instit}
\receipt{ 17 March 1995}
\begin{abstract}
 
In this work we study in detail the NeVII 2s3p-2s3s singlet line, which
 was also the object of a recent experiment. The standard perturbative impact
 theory  predictions are tested against a fully non-perturbative semiclassical
 impact calculation, taking into account dipole and quadrupole interactions.
 Potentially very significant  problems with the standard perturbative theory
 are encountered and discussed and a simple remedy is  proposed.

Published in Phys.Rev.Lett.{\bf 75}, 3406(1995). 
 
\end{abstract}
%{\tt$\backslash$\string pacs\{\}} should always be input,
%even if empty.}
 \narrowtext
%\widetext
 
%\section{    Introduction}
%\label{sec:level0}
 
The calculation of plasma-broadened
line spectra, provides a very useful
 diagnostic tool and additionally is a 
  necessary ingredient for large-scale
 computations in astrophysics and plasma physics.
  A  major cornerstone, reducing the many-body problem of line-broadening to
 the computation of one-body quantities, is the impact approximation
\cite{griem1,griem2}. For practical calculations,
 the impact theory is usually employed in its perturbative version, and even
 then simplified formulas are often used.
 
 Isolated lines\cite{griem2},
 by being relatively simple and usually unaffected by ion microfield 
 effects, whether static or dynamic\cite{SA1} are an excellent testing
 ground for the theories of electron collisional broadening. Such tests require
 reliable experimental profiles and much progress has been made in this
 direction in recent years
  mainly by the Bochum group\cite{Bot,GUK,GHK,Gl1,Uze,Neon,GK}.
 These studies have revealed significant discrepancies
 with simplified expressions that are often used for the electron
 collisional broadening\cite{griem2,GrSEM,DK,HB,GBK}. Furthermore, serious
 discrepancies with close-coupling(cc) 
 calculations\cite{Seaton}
 were found 
in\cite{GUK}. Even more impressive is a recently obtained factor of 2
 discrepancy between 
 cc calculations\cite{Seaton}
 and experiment\cite{GK}, for a line and parameter range where cc should
 be at its best. In both cases, much better agreement (roughly by a factor of
 2) is obtained by semiclassical(sc) calculations. This means that sc
 calculations are in fact $the$ $best$
 available today, in the sense of giving agreement with
 experiment.
 
 At the foundation of any  
  sophisticated\cite{SA1,GBKO,SA4,SB1,SB2,SA2}
  sc perturbative calculation
 is the requirement that unitarity is not violated. This is important, of
 course, since unitarity violation can lead to a serious overestimation
 of the width\cite{SA4,SA5}. Unitarity is preserved   
 for over 30 years  by
 using a criterion, thought to be both necessary and sufficient. This
  \lq 
  \lq fact" has gone unchallenged over this period of time. Hence, a
  minimum impact parameter   $ \rho_{min}(v)$  is determined, such that
 unitarity is satisfied for larger impact parameters,
  by numerically solving the equation
\begin{equation}
\vert
\lbrace{\sum_{a^\prime}\int_{-\infty}^\infty
V^\prime_{aa^\prime}
(t_1) dt_1  \int^{t_1}_{-\infty} dt_2 V^\prime_{a^\prime a}(t_2) +
\sum_{b^\prime} \int_{-\infty}^\infty
V^\prime_{bb^\prime}(t_1) dt_1 \int^{t_1}_{-\infty}
 dt_2 V^\prime_{b^\prime b}(t_2) \over\hbar^2}\rbrace 
\vert
\le d
 \end{equation}
where $\lbrace\ldots\rbrace$ denotes an angular average,
$V^\prime$ denotes the
 sc\cite{griem1}
 emitter-perturber interaction in the
 interaction picture, $a$ and $b$ denote upper and lower level
 states respectively, 
 $a^\prime$ and 
 $b^\prime$ denote a complete set of states that perturb
  $a$ and $b$ respectively,
 and
 where d is a number less
 than or equal to 1. The condition d=1 is sufficient to preserve
 unitarity, but  to keep the expansion parameter small, frequently d=0.5 is
 also used\cite{SB2}. Often, the real part of the LHS of the expression
 in eq.(1) is substituted for the absolute value sign, because the
 imaginary part is usually much smaller.  
 The test in eq.1 is then carried out using the A-function\cite{griem2,SA1,SB1}
  for which analytic expressions have
 been recently given\cite{SA3}. Thus, one hopes that perturbation theory is
 reliable for
 $ \rho \ge  \rho_{min}(v)
 $   and that this region gives the dominant contribution, so that one may
 either neglect or estimate very roughly the unknown (within a sc
 perturbative framework)  $ \rho \le  \rho_{min}(v)
$ contribution. In this work, we demonstrate by means of a $specific$
 $case$ that checking unitarity in this manner
 is not sufficient and the error associated with the breakdown of this
 test can be substantial.
 
The main problem of the impact approximation has always been the so-called
 strong collisions, i.e. the collisions at  small impact parameters
  $ \rho \le  \rho_{min}(v)$.
   It has been known for a long time that
the contribution of such collisions can be  bounded, but usually some
 fraction of this bound is then used as an additive contribution to the
 width. Whereas for say large scale computations, returning a single
 value for the width is desirable, for detailed comparisons with
 experimental results it is best to return an error bar for the contribution
 of these strong collisions, which cannot be computed reliably. This is
 the approach adopted here\cite{SA1}. 
  In other words the theoretical width lies between the weak collision
 width ($\rho\ge\rho_{min}(v)$), which is $presumed$ to have been
 reliably computed by the usual perturbative(PR) treatment, and the sum
 of the weak and strong ($\rho\le\rho_{min}(v)$) collision
 widths.
Sometimes these bounds were satisfactory\cite{SA1}, while other
times\cite{Gl1}
 they were rather too large and  reducing them  would be desirable.

Although one loosely speaks of small impact parameters and/or small velocities
 as giving rise to strong collisions, it is important to distinguish
 between collisions which fall into this group because unitarity is
 violated as the perturbation expansion fails
  and collisions which fall into this group because the
 sc approximation is no longer valid. Whereas there is no way
 within the sc approximation to avoid the later,  the
 collisions which violate unitarity are still treatable within the sc
 impact approximation, by means of a non-perturbative(NP) approach.
 There are essentially two such known NP approaches. The first
 one is an analytic solution  under the approximation of neglecting
 time-ordering effects, while the second\cite{Bacon} is a fully
 numerical solution
 of the Schr\"odinger equation. To avoid ambiguities, we have chosen the
 second approach here. Thus, if the
 $\rho_{min}(v)$  required to satisfy unitarity is significantly larger
 than  then 
 $\rho_{min}(v)$  required to preserve the sc approach,
 as is very often the  
 case\cite{GHK,Gl1,Uze,Neon,GK}
 a  NP
calculation achieves a $very$ $significant$ reduction of the error bars
  given by the PR calculation.
 
 As a reminder,
 the half width(HWHM) is written as\cite{SB1}
 \begin{equation}
  HWHM=2  \pi n \int \rho d\rho \int dv v f(v) Q( \rho,v)
  \end{equation}
    where n is the electron density,
 $f(v) $ is the Maxwelian velocity  distribution and Q is given by
 \begin{equation}
 Q(\rho,v)=\lbrace I-S_a(\rho ,v) S^{-1}_b(\rho ,v) \rbrace
   \end{equation}
  where the subscripts a and b denote the upper and lower levels
 respectively,
 I is the unit matrix, S is the S-matrix and
  $ \lbrace \ldots \rbrace$   denotes angular average.
    When one formulates the problem on the collision axes,
%(since  the sc perturber motion takes place in a plane),
 one finds for isolated lines\cite{SB1}
  \begin{eqnarray}
&&Q(\rho ,v)=(2J_a+1)^{-1}\sum_{M,m_a,m_a^
\prime,m_b,m_b^\prime}\langle J_b m_b^\prime 1M\vert J_a m_a^\prime\rangle
\nonumber\\ &&
 \langle J_b m_b 1M\vert J_am_a\rangle\lbrack
\delta_{m_b,m_b^\prime}\delta_{m_a,m_a^\prime}
\nonumber\\ &&-\langle J_a m_a^\prime
\vert S_c( \rho ,v)\vert J_a m_a\rangle
\langle J_b m_b^\prime\vert
S_c( \rho ,v) \vert J_b m_b\rangle^\star \rbrack
\end{eqnarray}
 where the subscript c denotes
 that the S-matrix has been computed for a given direction
 of the perturber trajectory ( collision axes).
 
For ion lines, the trajectory is parametrized in
terms of the "time" variable u, defined
 by \cite{SA1,SB1}
 \begin{equation}
t=s( \epsilon sinh
 \lbrack 
u 
 \rbrack 
-u)/v
 \end{equation}
 with   $s={ Z_{em} e^2  \over 4 \pi \epsilon_0 mv^2 }  $
and with the eccentricity

 All calculations described in this work refer to
 the singlet 2s3p-2s3s NeVII 3643.6\AA $\  $  line,
 which has been the subject
 of a recent experimental study\cite{Neon}. The combination of available
 experimental data and the fact that broadening is determined by a small
 set of three levels makes this line suitable
 for illustrating the important   points.
 We first give an example of a $pure$ $dipole$ calculation.
  When we calculate
  $Q(\rho ,v) $  for  a velocity of
 2x10$^6 $m/sec and an impact  parameter
 of 5\AA ($\epsilon=$ 1.653), the PR result is  $Q(\rho ,v)$
=  0.05108, and is in good agreement with the NP
 result of 0.04997. This is understandable, since
 $Q(\rho ,v)$ is small(=0.05108), so the perturbation
 expansion should be accurate. But when we attempt the same calculation for an
 impact parameter of 0.68\AA
 ($\epsilon$ =1.0158), which also has a relatively small  expansion
 parameter of $Q(\rho ,v)$ =0.16, the NP result is 0.0868.
 Fig.1, which shows the real part of
 $\lbrace I-U_a(u)U^{-1}_b(u)\rbrace $ versus the "time" variable u,
  illustrates what has happened: If the U-matrices remain close to the
 unit matrix, then the  perturbation expansion is certainely valid. This is the
 case of the solid line, which  represents the larger eccentricity.
 However,  we may get a small
 $Q(\rho ,v) $
    without this being the case, as is demonstrated by the dashed line, which
 represents the smaller eccentricity. In other words,  it is only
 for small times (or u) that the U-matrices evolve according to
 perturbation theory in the smaller $\epsilon$ case.
 Consequently, the unitarity criterion is insufficient and the  NP
 result is substantially different from the PR one.
  We note the flat initial and final regions for both cases in Fig.1,
 which show
 that the U-matrix has indeed converged to the S-matrix.
 
\underline{Fig.1 goes here}
 
Nevertheless, the PR and NP HWHM for a dipole-only
 interaction turn out to be close: 0.535 vs. 0.5094\AA $
\   $  respectively. The
 reason for this agreement can be seen in Fig.2, where the PR
 (solid line) and NP (dashed line) $Q(\rho ,v) $
   are shown as a function of $\rho $ for different velocities. Keeping in mind
 that  the average velocity is 2.9x10$^6$ m/sec, it is clear why  the
 differences between the PR and NP calculations are
 small: The main contribution comes from weak collisions for which the
 PR and NP calculations are in good agreement.
 
\underline{Fig.2 goes here}
 
Things are different in the quadrupole case where the dominant contribution
 comes from smaller impact parameters and velocities.
 In PR calculations, unitarity
 requires a cutoff at larger impact parameters compared to
 a pure dipole calculation, with  an associated, often
 very significant, increase in the strong collision error bars.
  In such calculations including
 quadrupole impact broadening often\cite{SB1} only the diagonal channels are
 included. In \cite{SA2} this assumption was discussed and a rough criterion
 was adopted for  including the "near hydrogenic" nondiagonal channels.
 Although tables of the relevant nondiagonal functions exist\cite{Klar},
analytic expressions have not been developped for these functions, unlike
 the dipole case, and it would be desirable to do so, in order to
  be able to determine which quadrupole channels should be included in the
 calculation.
  We here present  concrete examples where the "near hydrogenic"
 nondiagonal channels
 must be included: For the line in question, the  only diagonal quadrupole
channel is 3p-3p, since 3s-3s is not allowed. However, the 3s-3d channel is
 allowed and this is often neglected\cite{SB1}. Fig.3
  shows a comparison of
 pure quadrupole calculations. There are basically two regimes, the
 perturbative regime and the strong collision regime at small
 impact parameters. The first thing to note is that $even$ in the
 PR regime, $it$ $makes$ a significant difference whether the
 3s-3d channel is included or not. On the other hand, because
 perturbation theory is valid, the 3d-3d channel makes no difference
  in the results. This  changes in the strong collision regime
 and the addition of the 3d-3d channel makes an important difference,
 resulting in a peak rather than a trough.  This result confirms that
 as we move to smaller $\rho$, progressively more and more
  perturbing levels and channels come into play.
 
\underline{Fig. 3 goes here}
 
 Fig.4 is a similar graph, but with the dipole terms included.
 It should be noted  that  the strong collision regime shrinks  to
 smaller $\rho$ for a pure dipole interaction or if the velocity
 is increased. The remarks made about the importance of the
 \lq\lq nearly hydrogenic" 3s-3d channels still hold true.
 
\underline{Fig. 4 goes here}
 
 With regard to the effect on the final line profile,
  the electron temperature (average velocity) and density
 (average impact parameter) are very important in determining
 the relative significance of the strong collision regime. For the
 Neon line considered here\cite{Neon}, the strong collision
 regime is very important for a dipole plus
 quadrupole calculation:the PR result is 0.563\AA with an error
 bar of 0.477$\AA$,
  as opposed to error  bars of about 0.13$\AA$ for the NP
 results, for which the calculations involving the a) 3p-3p,b) 3p-3p and
 3s-3d, and c) all 3p-3p,3s-3d and 3d-3d
 quadrupole channels  yield HWHMs of 0.85, 0.977 and
 0.975\AA$ \ $  respectively.
(The experimental result 
  is \cite{Neon}
   0.865
 \AA.)
It is noteworthy that  the 3s-3d channel accounts for about 15\%
 of the $total$ (i.e. including the dipole interaction)
 electronic width, whereas the 3d-3d channel results in a decrease of
 the $total$ electronic width by less than 0.5\%. These results
 thus demonstrate the need for analytical approximations of the
 relevant nonhydrogenic quadrupole broadening functions\cite{Klar},
 which should then be included in existing codes. These results(with
 their error bars) are absolutely
 rigorous(within the number of states used), if one trusts the  
 \lq\lq demarkation line" between what may and what may not be treated
 semiclassically, and for which the convention
 in\cite{SA1} was adopted. It is then interesting that even for this
 case, where relatively low partial waves are involved, i.e. under conditions
 not most favorable for sc, good agreement is obtained.
 Furthermore, the effects desribed here were $responsible$ $for$
 the seemingly
 excess broadening observed in 
  NeVIII\cite{GUK}
 for which NP calculations give an electron FWHM contribution between 
  0.9  and 0.9+0.13
 \AA, with the later number corresponding to the (unlikely) maximum
 possible contribution from nonsemiclassical collisions, compared with
 an experimental result of 1.2$
 \pm 0.1$
 \AA at the highest density considered.
 
In summary, we have demonstrated that the standard practice of even the
 most sophisticated current  perturbative
 impact theory calculations, cannot enforce unitarity 
 reliably and that a more reliable  verification
 of the unitarity is essential. Such a procedure would involve
  additional information and could be implemented by evaluating, at
  intermediate times, the function 
  ${\lbrace I-U_aU_b^{-1}\rbrace}$ to ensure that it is monotonic.
  One possible implementation of this  could involve
 using an upper limit of the $t_1$ integrations of 0 rather than
 $\infty$ in Eq.(1); this check would allow the diagnosis of the
 situation shown by the dashed line in Fig.1.
 Even without any  additional checks, it is still much better
 to check than not to, since
 for large impact parameters, the unitarity check $is$  reliable.

 Further, it has also been demonstrated 
 that to the extent that quadrupole interactions
 are important, it is not always sufficient to include only the diagonal
 quadrupole channels. An analytic solution to the problem of strong
  collisions is the subject of current research.

\section{Figure Captions}

FIG.1 Real part of $\lbrace I-U_aU_b^{-1}\rbrace$ vs. u.
 The solid and dashed lines correspond to a NP calculation for
 $v=2x10^6$m/sec and $\rho=5$ and $\rho=0.68\AA$ respectively.
 
FIG.2 $Q(\rho,v)$ vs. $\rho$ for different velocities $v$. The
 solid  and dashed  lines represent respectively
PR and NP dipole calculations.
 
FIG.3 $Q(\rho,v)$ vs. $\rho$ for a pure quadrupole interaction. a)$v$=
2x10$^6$m/sec, b)$ v$=3x10$^6$m/sec. The solid, dotted and dashed lines
 correspond respectively to NP calculations including all, all except the
 3d-3d, and only the 3p-3p channels
 respectively. The dash-dotted line corresponds to
  a PR calculation including only the 3p-3p channel.

FIG.4 $Q(\rho,v)$ vs. $\rho$ for dipole and quadrupole interactions. a)
$v$=2x10$^6$m/sec, b) $v$=3x10$^6$m/sec. The solid, dotted and dashed lines
 correspond respectively to NP calculations including all dipole
 channels and, respectively, all,  the 3p-3p and 3s-3d,
  and the 3p-3p only
 quadrupole channels. The dash-dotted line is a pure dipole
 NP calculation.
\enddocument